%% file: main.tex
\documentclass[sigconf]{acmart}

\usepackage[capitalize, noabbrev]{cleveref}
\usepackage{listings}
\usepackage{subcaption}

\AtBeginDocument{%
  \providecommand\BibTeX{{%
    \normalfont B\kern-0.5em{\scshape i\kern-0.25em b}\kern-0.8em\TeX}}}

\copyrightyear{2024}
\acmYear{2024}
\setcopyright{rightsretained}
\acmConference[CHI '24]{Proceedings of the CHI Conference on Human Factors
in Computing Systems}{May 11--16, 2024}{Honolulu, HI, USA}
\acmBooktitle{Proceedings of the CHI Conference on Human Factors in
Computing Systems (CHI '24), May 11--16, 2024, Honolulu, HI, USA}
\acmDOI{10.1145/3613904.3642406}
\acmISBN{979-8-4007-0330-0/24/05}

\include{util}

\begin{document}

\title[Writer-Defined AI Personas for On-Demand Feedback Generation]{Writer-Defined AI Personas for On-Demand Feedback Generation}

\author{Karim Benharrak}
\orcid{0009-0002-3279-5664}
\email{karim@benharrak.com}
\affiliation{%
  \institution{The University of Texas at Austin}
  \city{Austin}
  \state{TX}
  \country{USA}
}
\affiliation{%
  \institution{University of Bayreuth}
  \streetaddress{Universitätsstr. 30}
  \city{Bayreuth}
  \state{Bavaria}
  \country{Germany}
  \postcode{95447}
}

\author{Tim Zindulka}
\orcid{0009-0009-1972-351X}
\email{tim.zindulka@uni-bayreuth.de}
\affiliation{%
  \institution{University of Bayreuth}
  \streetaddress{Universitätsstr. 30}
  \city{Bayreuth}
  \state{Bavaria}
  \country{Germany}
  \postcode{95447}
}

\author{Florian Lehmann}
\orcid{0000-0003-0201-867X}
\email{florian.lehmann@uni-bayreuth.de}
\affiliation{%
  \institution{University of Bayreuth}
  \streetaddress{Universitätsstr. 30}
  \city{Bayreuth}
  \state{Bavaria}
  \country{Germany}
  \postcode{95447}
}

\author{Hendrik Heuer}
\orcid{0000-0003-1919-9016}
\email{hheuer@uni-bremen.de}
\affiliation{%
  \institution{University of Bremen}
  \city{Bremen}
  \state{Bremen}
  \country{Germany}
  \postcode{28359}
}

\author{Daniel Buschek}
\orcid{0000-0002-0013-715X}
\email{daniel.buschek@uni-bayreuth.de}
\affiliation{%
  \institution{University of Bayreuth}
  \streetaddress{Universitätsstr. 30}
  \city{Bayreuth}
  \state{Bavaria}
  \country{Germany}
  \postcode{95447}
}

\renewcommand{\shortauthors}{Benharrak et al.}

\begin{abstract}

Compelling writing is tailored to its audience. 
This is challenging, as writers may struggle to empathize with readers, get feedback in time, or gain access to the target group. 
We propose a concept that generates on-demand feedback, based on writer-defined AI personas of any target audience. 
We explore this concept with a prototype (using GPT-3.5) in two user studies (N=5 and N=11): 
Writers appreciated the concept and strategically used personas for getting different perspectives. The feedback was seen as helpful and inspired revisions of text and personas, although it was often verbose and unspecific.
We discuss the impact of on-demand feedback, the limited representativity of contemporary AI systems, and further ideas for defining AI personas. 
This work contributes to the vision of supporting writers with AI by expanding the socio-technical perspective in AI tool design: To empower creators, we also need to keep in mind their relationship to an audience.

\end{abstract}




\begin{teaserfigure}
  \centering
  \includegraphics[width=\textwidth]{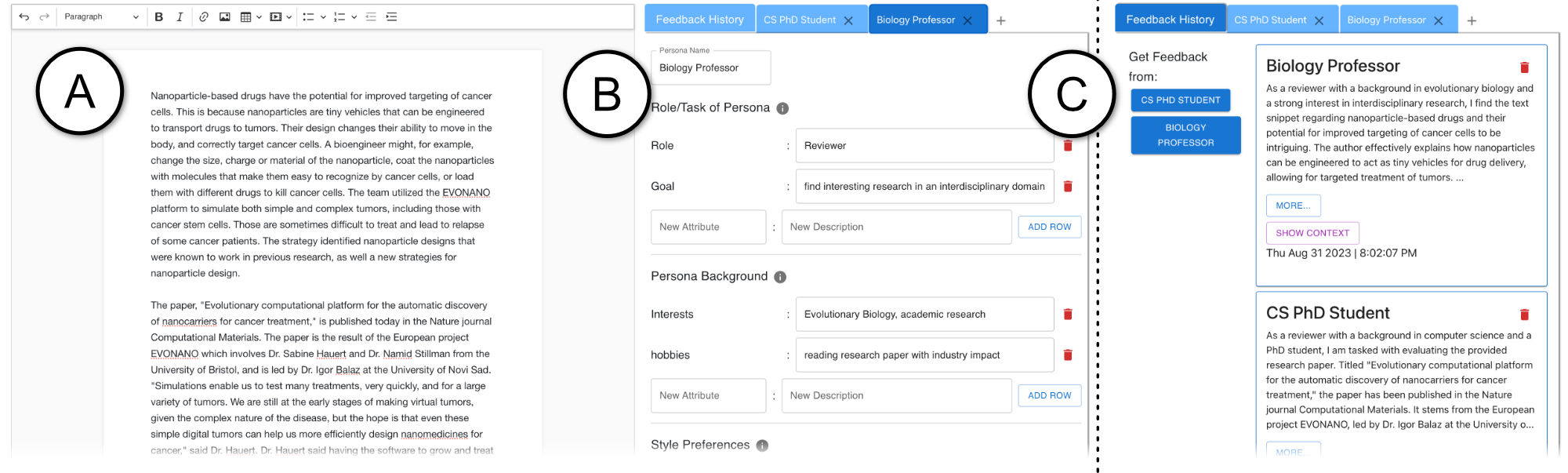}
  \caption{Our text editor \textit{\system} supports writers with generated feedback on their draft: While working on a text (A), writers can define \aipersonas{} that represent their target readers (B). Writers can select any part of their text and click on a persona button (C) to receive feedback from the perspective of that persona. This feedback is generated with a Large Language Model (GPT-3.5) by prompting it with the provided persona information. (B) and (C) show different tab states of the same sidebar, not two sidebars.}
  \label{fig:teaser}
  \Description{This figure is divided into three parts: A, B, and C. Part A shows a text editor. Part B shows a sidebar with Navigation Tabs "Feedback History", "CS PhD Student", and "Biology Professor" at the top. The "Biology Professor" tab is active. Below the navigation panel, there is a text field for "Persona's name" that is currently filled with "Biology Professor". There are three sections that can be seen ("Role/Task of Persona", "Persona Background", and "Style Preferences"), each having different rows of text field pairs with one extra row of empty text fields and a button "Add Row". The first section currently has the pairs "Rule": "Reviewer", "Goal": "find interesting research in an interdisciplinary domain". The second section consists of the pairs "Interests": "Evolutionary Biology, academic research" and "hobbies": "reading research paper with industry impact". Each row has a red trash icon on the right side. Part C shows a sidebar with the "Feedback History" tab being active. On the left side there is a headline "Get feedback from:" with buttons ("CS PhD Student", "Biology Professor") below. On the right side are two cards with separate headlines "Biology Professor" and "CS PhD Student". Each card contains written text (feedback) and different buttons: a red trash can button, "MORE...", and "SHOW CONTEXT". At the bottom of each card is a timestamp.}
\end{teaserfigure}

\maketitle

\input{content/01-intro.tex}

\input{content/02-background.tex}

\input{content/03-concept.tex}

\input{content/04-method.tex}

\input{content/05-evaluation.tex}

\input{content/06-discussion.tex}

\begin{acks}
Daniel Buschek is supported by a Google Research Scholar Award. This project is funded by the Bavarian State Ministry of Science and the Arts and coordinated by the Bavarian Research Institute for Digital Transformation (bidt).
\end{acks}

\bibliographystyle{ACM-Reference-Format}
\bibliography{references}

\appendix
\input{content/07-appendix.tex}

\end{document}

%% file: util.tex
\definecolor{brown}{rgb}{0.59, 0.29, 0.0}
\definecolor{darkgray}{rgb}{0.59, 0.59, 0.59}
\definecolor{tablegray}{gray}{.9}

\newcommand\revision[1]{#1}

\usepackage[utf8]{inputenc}
\usepackage{diagbox}
\usepackage{colortbl}

\usepackage{tabularx}

\usepackage{color}

\usepackage{enumitem}
\usepackage{mathtools}
\usepackage{commath}

\usepackage{calc}

\usepackage{amssymb}%
\usepackage{pifont}%

\newcommand{\customtilde}{{\raise.17ex\hbox{$\scriptstyle\sim$}}}

\usepackage{xparse}

\newcommand{\system}{Impressona}
\newcommand{\aipersonas}{AI personas}

\newcommand{\p}[1]{$P_{#1}$}

%% file: content/01-intro.tex
\section{Introduction}

People write \textit{for} people~\cite{hayes_modeling_2012}. Doing so well is challenging, as many readers of this paper will have experienced as writers themselves: Internal ideas need to be translated into a sequence of words to be brought to the page for the readers~\cite{flower1981cognitive}. For instance, researchers of any field need to convey their findings to their academic community, including peers and reviewers, but also policy makers and the general public, in reports, magazines, blogs, and so on. Here, writing excels if it conveys information in such a way that it is easily understandable and relatable to the target group.

Despite this social nature of writing~\cite{sturm2016schreibkompetenz}, writers are frequently left disconnected from their readers. %
Existing writing tools address this only to a limited extent: %
For example, the text checking tool \textit{Grammarly} includes an ``audience'' setting (three levels of expertise)~\cite{Grammarly2020goals}. Already in 2010, \citet{Bernstein2010soylent} proposed a different approach with \textit{Soylent}, a text editor that allowed writers to delegate tasks to crowdworkers. More recently, systems such as \textit{Wordcraft}~\cite{yuan_wordcraft_2022} or \textit{CoAuthor}~\cite{lee_coauthor_2022} provide co-writing experiences with AI. None of these approaches bring in the perspective of the target readers.

Doing so is challenging: It can be difficult for writers to get access to a target group during writing sessions, in time in between sessions, or at all. Thus, writers are rarely able to receive feedback ``on-demand'' from a reader perspective. %

The goal of this paper is to address this and support writers in receiving feedback from the perspective of their target group through technology.
As our design inspiration, we note parallels to user-centred design. Just as writers write for a specific group of readers, designers create a digital artefact for a target group. This target group may simply live in the back of the designer's mind but it is often beneficial to engage with it more concretely, for instance, via user research and the creation of personas. 

We transfer this to writing and propose the concept of \textit{\aipersonas{} for feedback}: Writers define personas of their target readers, which then provide feedback on the text. This feedback is generated by a Large Language Model (LLM) based on a prompt that includes the persona information. We investigate this concept with three research questions:

\begin{itemize}
    \item \textbf{RQ1:} How do writers perceive the idea of AI personas?
    \item \textbf{RQ2:} How do writers define and use AI personas?
    \item \textbf{RQ3:} How well can contemporary LLMs support AI personas?
\end{itemize}

To address these questions, we developed \textit{\system}, the first text editor that enables writers to create \aipersonas{} of their target readers to receive feedback (\cref{fig:teaser}). 
While working on a text (\cref{fig:teaser}A), writers first create a persona and enter its attributes (\cref{fig:teaser}B). Examples for such attributes include the persona's background knowledge (e.g. ``computer science basics''), job (e.g. ``professor'') and role (e.g. ``editor''). %
They can create multiple personas and thus reader perspectives. Writers select parts of their text and click on a persona to generate comments (\cref{fig:teaser}C). Our backend prompts OpenAI's GPT-3.5 to generate feedback, with the selected text and the persona information embedded into a prompt template.
For an example of one such interaction moment from our user study, see \cref{sec:appendix_interaction}.

We use our prototype in two user studies (N=5 and N=11), with a design iteration in between. Participants brought their own writing projects to our study to work on real-world texts that are personally relevant to them. %
They appreciated the on-demand concept and strategically used personas for getting different perspectives. However, they found it challenging to initially define personas. While the feedback was often verbose and unspecific, it was seen as helpful and inspired revisions of text and personas.

In summary, we contribute: (1) A novel concept for writers to receive feedback, generated based on writer-defined AI personas that represent target readers, (2) its implementation in a functional prototype, and (3) insights from an evaluation with users.

In a broader view, we contribute to the vision of supporting writers with AI by expanding the socio-technical perspective in AI tool design: To support creators, we also need to keep in mind their relationship to an audience.

%% file: content/02-background.tex
\section{Background \& Related Work}
\label{sec:related_work}
We clarify similarities and differences between personas in user-centred design and our concept of \aipersonas. We further relate our work to research on writing and writing support.

\subsection{From Personas in UCD to AI Personas of Readers}

A persona in user-centred design (UCD)~\cite{Cooper2014AboutFace} is a descriptive model of a group of users that captures information about how they behave and think, and what their motivations, needs and goals are. %
They help designers to avoid ``self-referential design''. Analogously, our goal is to help writers empathise with their readers and avoid ``writer-based prose'', which stays in the writer's own information organisation that might not be useful for readers~\cite{flower1981cognitive}. %

The key difference to UCD is that writer-defined personas are likely not informed by user research. We also do not expect writers to be familiar with the concept of personas in UCD. %
Their ``personas'' of readers may nevertheless be grounded in experience -- e.g. from past feedback by readers, by genre conventions, and so on. Recent HCI research suggests that writers indeed develop a mental model of whom (or which system) to approach for what kind of feedback~\cite{Gero2023}. Similarly, writing research discussed the concept of a ``mental convention model''~\cite{gopferich2009comprehensibility} that writers internalise as they become familiar with a writing context (e.g. genre, target community). These findings lend credence to the idea that writers can describe aspects of their target readers to generate feedback on.

\subsection{Using Large Language Models for the Representation of People}
\label{sec:background_nlp}

Related work leveraged Natural Language Processing (NLP) to represent or ``simulate'' people~\cite{Park2023generative}. A key line of work refers to ``personas'' as a foundation for generating text with language models from a certain perspective (e.g.~\cite{Song2019exploitingpersona, Song2021bobpersona, zhang2018personalizing, Zheng2020pretrainingpersona}), with HCI applications particularly for chatbots (e.g.~\cite{6569323, BanterBotMediumArticle, Schmitt2021, sun2023inspire}). These personas are often short text descriptions. A related concept is the ``system instruction'' prompt, which tells an (instruction-tuned) model which perspective to take (e.g. ``You are an assistant that speaks like Shakespeare''\footnote{\url{https://help.openai.com/en/articles/7042661-chatgpt-api-transition-guide}}). 

Related to writing, \citet{nilsson2016dead} introduced the concept of semantic avatars of (historic) authors based on a language model. %
More recently, \citet{10.1145/3544548.3580688} used an LLM to generate synthetic responses for HCI experiments. Similarly, \citet{Park2022socialsimulacra} used LLMs to generate user content during prototyping, such as comments for a discussion board. They included short text statements as ``user personas'' in their prompts (e.g. name + job). %

While systems based on LLMs have received attention for their ability to generate human-like language~\cite{roose2022brilliance},  %
researchers have also pointed out their limitations. For instance, \citet{kabir2023answers} %
find that more than half of ChatGPT's responses to software engineering queries contain inaccuracies~(52\%). %

These potentials and limitations motivated us to explore how LLM-based AI systems can support writers in practice, specifically through the lens of reader personas. %

\subsection{Writing Research \revision{\& Practice}: Writing for an Audience}
\label{sec:background_writing_guides}

Thinking about the target readers is common advice in writing guides. \citet{rechenberg_technisches_2006} motivates this as a safeguard against overly abstract writing. Moreover, he highlights that writers should continuously ``talk'' to an imagined inner reader to help them think about their readers' existing knowledge and known terminology, and anticipate questions. 
Similarly, \citet{langer2019verstandlich} describe the concept of ``person-centred writing''. One of its key aspects is \textit{empathy} -- striving to imagine the readers' world of experience.
We aim to support writers in doing so, by helping them to externalise the inner reader as a persona and by using AI to let it ``talk''.

Related, \citet{gopferich2009comprehensibility} highlights the ``communicative function'' of texts in her framework on evaluating comprehensibility. One of the framework's dimensions is the ``target group'', with features such as age, gender, cultural background, education, hobbies, and prior knowledge, as well as the heterogeneity of the group(s). In our prototype, writers can describe reader personas along such dimensions, and create multiple different personas.

If writers should think about an inner reader, when do they think about the text? The different cognitive processes in writing are non-linear, as described by Flower and Hayes~\cite{Hayes1980, flower1981cognitive}, involving planning, translating, and reviewing. %
Writers consider the target audience on several levels of planning, from concrete content goals and plans (e.g. what to say/do to an audience), to middle-range goals (e.g. ``appealing to an audience''), to top-level goals (e.g. ``interest the reader''). Writers work on ideas and plans towards these goals through a pattern of exploration and consolidation. %
Our fundamental UI design accounts for this: Writers can easily switch from writing in a \textit{page view} to a \textit{sidebar} for capturing their understanding of the audience by creating and editing personas. %

\subsection{Designing Writing Tools: Expanding the Perspective Beyond the Writer}
\label{sec:perspective_beyond_writer}

Writing tools have a long history~\cite{macdonald1982writer, Kirschenbaum2016}. %
NLP capabilities are increasingly integrated into them, both in industry (e.g. ``Copilot'' in MS Office~\cite{MicrosoftBlogCopilot}), as well as in the HCI research community and its prototypes (e.g. see~\cite{Chang2023in2writing, Gero2022designspace, Huang2022in2writing, lee2024dsiiwa}). %

A key observation here is that most writing tools share a common perspective on the socio-technical environment that is relevant to their design: There is one person writing, to be supported by technology. As recently shown, acceptance of such support is impacted by social dynamics~\cite{Gero2023}. Related, AI-mediated communication~\cite{Hancock2020JCMC} examines sender and receiver experiences of text written with AI (e.g.~\cite{Hohenstein2020moralcrumple, Jakesch2019aimediated, Mieczkowski2021cscw}). Together, these observations motivate us to explore a design perspective that considers that most writing serves to communicate information to others. %

A specific related example is provided by recent research into writing support for experts for plain and simple language~\cite{10.1145/3603555.3603569,10.1145/3544549.3585749}: These specialist writers found it important to have their text evaluated by stakeholders and felt that technology could support such reviews. We provide a first step towards supporting such feedback loops with AI.

\subsection{The Role of AI Tools and Feedback from Other People when Writing}
\label{sec:background_writing_feedback}

Writing support can come not only from AI tools, but also from peers. \citet{Gero2023} interviewed creative writers about both options. They report that writers ``learn specific characteristics of people that would modulate when they’d turn to these people for support.'' They also discuss individuality: AI reflects assumptions and a perspective based on -- and biased by -- its training data, which is different from an individual human. In this light, they call for ``[m]aking explicit these assumptions, as well as highlighting when a model is trained to produce a different perspective''. 
Related, recent work showed that a biased language model can influence viewpoints expressed by writers~\cite{10.1145/3544548.3581196}. This is particularly noteworthy considering concerns about which human values are represented by LLMs~\cite{johnson2022ghost}. 
In this light, we explore how writers might be \textit{supported} by perspectives provided by a (pretrained) LLM, when shaped through prompts based on the writer-defined persona information. %

%% file: content/03-concept.tex
\section{Concept and Implementation}
\label{sec:system}

We introduce the concept of \aipersonas{} of readers, a novel approach inspired by combining personas from user-centered design with insights from research into writing and AI writing support systems (see \cref{sec:related_work}). 
To explore this concept, we developed a prototype text editor called \textit{\system} (\cref{fig:teaser}).

\subsection{Concept Development}

We developed the concept and prototype based on insights from the literature and a design iteration with feedback from users.

\subsubsection{Design Iteration}
\label{sec:design_iteration}

We developed the prototype in two iterations.
First, we implemented an initial prototype (V1) and tested it with five participants.
Findings from this first study round motivated these three UI changes for V2:

\begin{itemize}
    \item \textit{Persona definition:} V1 had a single field for free text input. Our findings motivated more guidance here. Thus, we changed this into a form-like view in V2, to give it structure (attribute-description pairs). We also added an info button that shows examples.
    \item \textit{Multiple personas:} One persona at a time was not enough. We therefore changed the prototype in V2 to allow for creating multiple personas and switching between them via a tabs view.
    \item \textit{Feedback history:} Participants in the first study wished to review previously generated feedback, but V1 only showed the last feedback. This motivated us to integrate the ``Feedback History'' in V2.
\end{itemize}

In the rest of the paper, we always refer to the second version (V2), if not stated otherwise as V1.

\subsubsection{Design Rationale \revision{and Goals}}\label{sec:design_goals}

\revision{We extracted the following higher-level design rationale and goals from the literature and insights from our first user test with V1: 1) As the key idea, \textit{supporting on-demand feedback} allows writers to receive feedback from a reader perspective while writing, through AI personas. 2) \textit{Enabling writers to define personas themselves} allows them to express their target readers' characteristics, and to obtain different perspectives in this way. 3) Related, \textit{supporting easy iteration of personas} enables writers to refine them and also to react to potentially unwanted aspects in the generated feedback. 4) \textit{Supporting the definition and comparison of multiple personas} facilitates gaining insights through contrast, both into the draft, as well as into the AI's capability to manifest diversity in its generations. 5) Finally, \textit{supporting local feedback} allows writers to request feedback on any piece of text (which we realized through text selection) to facilitate specificity of feedback.}

\subsection{Frontend}
The frontend of our system is implemented as a web application using React \revision{to achieve the mentioned design goals}. %

\subsubsection{Main layout: Editor and sidebar}
The UI is split into two main areas -- a text editor (\cref{fig:teaser}A) %
and a sidebar (\cref{fig:teaser}B). %
The sidebar's navigation panel at the top consists of a ``Feedback History'' tab for requesting feedback, one tab for each created persona (to edit it), and a ``+'' button for adding a new persona. 
\revision{This side-by-side layout supports the mentioned design goals via easy access to both the draft and the personas.}

\subsubsection{Persona tabs: Create and edit personas}
\revision{This tab view supports the two design goals of enabling writers to define personas themselves and to easily iterate on personas.} Each persona tab leads to a view with form fields (\cref{fig:teaser}B). Writers can give their persona a name that also shows up in relevant other places (tab header, feedback by that persona). The rest of the form is structured in rows of pairs of text fields (for attributes and descriptions, e.g. ``job: physician''). Rows can be added and removed by clicking on respective buttons. %

\begin{figure}[t]
    \centering
    \includegraphics[width=\minof{\columnwidth}{0.6\textwidth}]{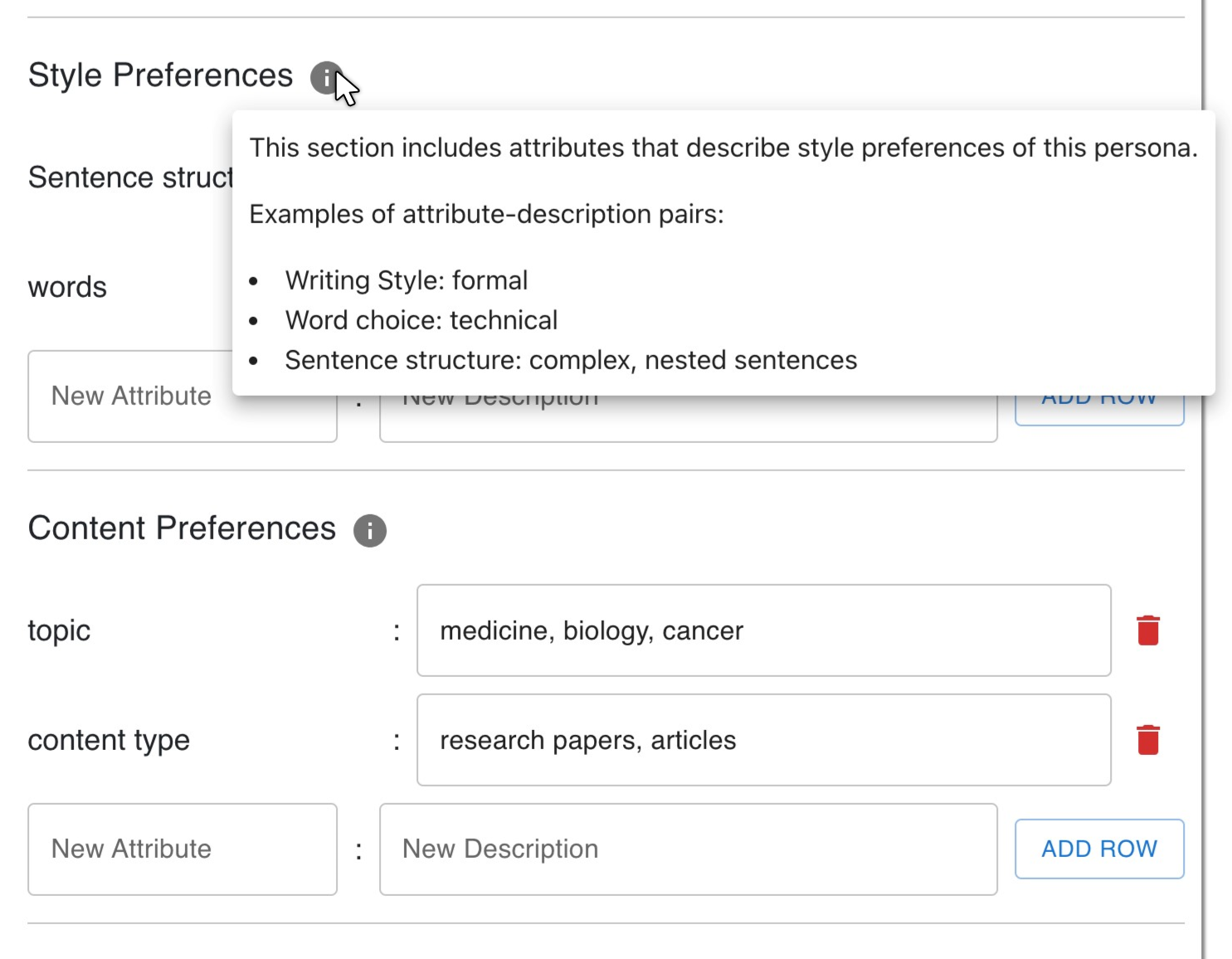}
    \caption{A part of the UI for defining a persona: Hovering over the info button next to each category (e.g. ``Style Preferences'') provides a corresponding description, alongside examples of attribute-description pairs to facilitate the persona creation process.}
    \label{fig:guidance}
    \Description{This figure shows a screenshot of the sidebar. At the top is the headline "Style Preferences" with an info button right next to it. The mouse cursor is hovering on the info button, which opened a pop-up that provides guidance: "This section includes attributes that describe style preferences of this persona. Examples of attribute-description pairs: Writing Style: formal, Word choice: technical, Sentence structure: complex, nested sentences".}
\end{figure}

This form has four sections: ``Role/Task of Persona'', ``Persona Background'', ``Style Preferences'', and ``Content Preferences''. %
An info button in each section header provides guidance (\cref{fig:guidance}). Clicking on these info buttons presents users with a short description and a set of examples of attribute-description pairs that could be added to this section. Writers can choose what to use -- it is not required to add information to all categories.

\subsubsection{Feedback tab: View and request feedback from the personas}
The ``Feedback History'' tab (\cref{fig:teaser}C) \revision{realizes the main design goal of providing on-demand feedback, as well as the goal of allowing for comparisons of feedback from multiple personas. It} holds a list of interactive cards that display the generated feedback. New feedback is displayed at the top. Cards start in a collapsed state that cuts off the text after a few sentences. A ``See more'' button allows users to expand the card to view the full feedback. A ``Show context'' button at the bottom of the card can be clicked to show the part of the text that the feedback was based on.

\subsubsection{Interaction}

\begin{figure*}[t]
    \centering
    \includegraphics[width=\linewidth]{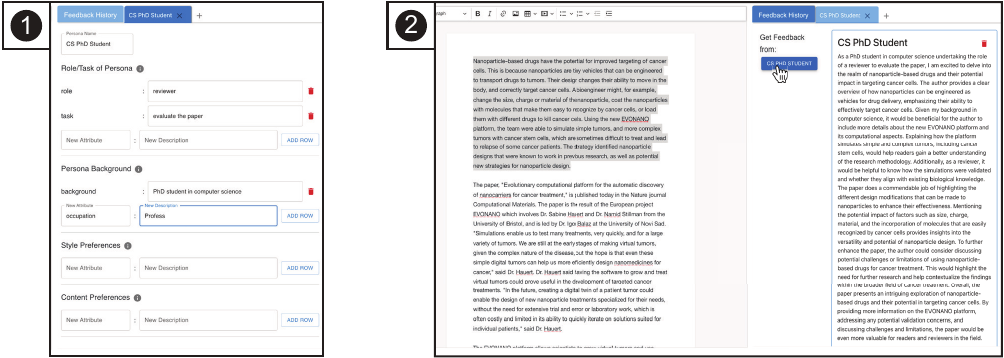}
    \caption{The overall workflow: (1) The writer defines a persona by adding rows of attributes and descriptions in a persona tab. (2) In the text editor, the writer selects a part of the text and then clicks on a persona button in the sidebar to generate feedback from that persona for that piece of text.}
    \label{fig:overall_workflow}
    \Description{This figure consists of two parts labeled "1" and "2". Part 1 shows the sidebar with a persona tab being active. Part 2 shows the editor and the sidebar with the "Feedback History" tab being active. The mouse cursor hovers above one of the buttons to initiate generating feedback from a persona.}
\end{figure*}

We realized three interactions with the personas: %

\paragraph{Persona creation} 
Writers click a ``+'' button in the sidebar's navigation panel to create a new persona. They then define their persona by filling out the form fields in its tab. %

\paragraph{On-demand feedback} 
To request feedback from a created persona, writers first select the part of their text they want to receive feedback on \revision{(in line with the design goal of supporting local feedback)}. They then click on a persona's button in the ``Feedback History'' tab. This adds a new comment card in the feedback history, displaying the feedback.

\paragraph{Persona refinement}
If writers are not satisfied by a response, or identified extensions or new directions for the persona, %
they can switch to the persona's tab at any time, to add, delete or edit fields; \revision{this supports easy iteration}. %

\subsection{Backend}
Our backend serves as a layer between our frontend and the OpenAI API\footnote{\url{https://platform.openai.com/}} that we use to generate the feedback.

\subsubsection{Implementation}
We implemented our backend as a Python Flask API. Our backend receives the selected text and the clicked persona from the frontend. We use this information to create a prompt that we send to the OpenAI API to generate the feedback. Upon receiving the response, the backend sends it to the frontend, to be displayed in a feedback card.

\subsubsection{Feedback Generation}
We leverage OpenAI's Chat Completion using GPT-3.5 to generate the persona feedback. See \cref{sec:appendix_prompts} for our prompt template.
We provided six input-output examples with this prompt template, as few-shot examples for the model. This was based on our observations in early experiments to improve the model's adherence to the desired output.
The feedback in these examples connected persona attributes to specific suggestions, to reveal how a persona's attributes affect the generated feedback (e.g. ``As a history professor, I value that this text...'').
These six examples were created from our own interactions with the OpenAI model, based on interactive sessions in which we explored the model's suitability at the start of the project (i.e. the few-shot examples are ``cherry-picked'' and manually refined input-outputs).  %

%% file: content/04-method.tex
\section{Method}
\label{sec:method}
We conducted two study rounds with our prototype, with a design iteration in between (\cref{sec:design_iteration}). Both studies used the same methods -- a writing task with think-aloud, a semi-structured interview, and a final questionnaire.

\subsection{Apparatus}
We used the prototype as described in \cref{sec:system}, hosted as a web app: The first study used V1, the second study used V2. Besides the features described earlier, V2 also implemented functionality to record user interactions with the system (e.g. creating personas, changing them, requesting feedback).

\subsection{Participants}
\label{sec:participants}
In total, 16 people participated in our evaluation (\revision{\cref{tab:participants}}): five in the first study (2 male, 3 female; referenced as \p{1} - \p{5}), using prototype V1; and eleven in the second one (7 male, 4 female; \p{6} - \p{16}), using V2. Their age ranged from 22-33 years. All were proficient English speakers; four were native speakers.

We recruited participants from our networks across a few universities. Our goal was to find people with real writing projects. In the first study, we were open to include various kinds of texts, to explore the concept. In the second study, we focused on the academic context, as one relevant target group for new writing tools (cf.~\cite{Strobl2019}). We report the resulting text types in \cref{sec:results_rq1}. %
Participants' backgrounds included Master students and PhD students as well as researchers.
Their median self-reported writing proficiency was 4 on a scale from (1) low proficiency to (5) high proficiency. %
The audiences participants regularly write for include \textit{scientific community, myself, professors, general public, managers, and colleagues}.
Participants in our second study were generally interested in topics around Artificial Intelligence (see \cref{fig:likert_results}).
Participants were compensated with €\,15.

\begin{table*}[t]
\footnotesize
\newcolumntype{L}{>{\raggedright\arraybackslash}X}
\newcolumntype{P}[1]{>{\raggedright\arraybackslash}p{#1}}
\newcolumntype{C}[1]{>{\arraybackslash\centering}p{#1}}
\renewcommand{\arraystretch}{1.5}
\begin{tabularx}{\textwidth}{P{.5cm} P{3cm} L C{1cm} C{1cm} P{.5cm} P{1cm} P{2cm}}
\toprule
\textbf{P} & \textbf{Writing Project} & \multicolumn{1}{l}{\textbf{Regular Audiences}} & \multicolumn{1}{l}{\textbf{Writing*}} & \textbf{English**} & \multicolumn{1}{l}{\textbf{Age}} & \textbf{Gender} & \textbf{Occupation} \\
\midrule
P1 & Bachelor thesis, Email to Professor, Fiction Story & Creative Writing, Video Game Writing, Scientific Writing & 4 & 5 & 25 & Female & Student \\
P2 & Blog post (AI in education) & P2 themselves, professors & 4 & 5 & 26 & Female & Student \\
P3 & Research Paper (HCI) & Scientific community & 4 & 5 & 32 & Male & Researcher \\
P4 & Research Grant Proposal (Computer Science) & Researchers, general public, students & 4 & 4 & 33 & Male & Associate Professor \\
P5 & Blog post (Cooking) & P5 themselves, professors & 3 & 3 & 32 & Female & Student \\
P6 & Web article (Computer Architecture) & Academic audience & 5 & 5 & 23 & Male & Student \\
P7 & Business Report (ERP system in HR) & Managers, Co-workers, professors & 2 & 4 & 22 & Male & IT-Consultant \\
P8 & Research Paper (HCI) & Scientific reviewers, Management Teams, C-Level of coporates/start-ups & 4 & 4 & 29 & Female & PhD equivalent \\
P9 & Research Paper (HCI) & Human-Computer Interaction and Human-Robot Interaction audiences & 4 & 5 & 33 & Male & PhD equivalent \\
P10 & Research Paper (HCI) & Domain experts, researchers, colleagues & 5 & 5 & 32 & Male & PhD equivalent \\
P11 & Research Paper (Process Analytics) & Computer Science scientists & 3 & 5 & 27 & Male & PhD equivalent \\
P12 & Research Paper (HCI) & Scientific conferences, academia, work-related audiences & 5 & 5 & 25 & Male & PhD equivalent \\
P13 & Research Paper (NLP) & Academic audiences in the fields of NLP, Machine Learning, etc. & 4 & 5 & 23 & Male & PhD equivalent \\
P14 & Masters Thesis Abstract (AI in medicine) & None & 2 & 3 & 27 & Female & Patient management in hospital \\
P15 & Research Paper Summary (Canine psychology) & Uni (exams, course work, presentations) & 3 & 5 & 31 & Female & Student \\
P16 & Research Essay (English language variety) & Professors at university & 3 & 5 & 26 & Female & Student \\
\bottomrule

\end{tabularx}
\caption{\revision{Overview of the participants. *Writing proficiency, **English proficiency (both on Likert scales, 1/low to 5/high).}}
\Description{This table gives an overview of the participants including columns that describe the writing project, regular audiences, writing proficiency, English proficiency, age, gender, and occupation for each participant.}
\label{tab:participants}
\end{table*}

\subsection{Procedure}
\label{sec:procedure}
\begin{figure}
    \centering
    \includegraphics[width=\minof{\columnwidth}{0.6\textwidth}]{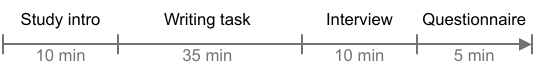}
    \caption{The study procedure, as described in \cref{sec:procedure}.}
    \label{fig:study_procedure}
    \Description{This figure shows a timeline. The timeline is divided into four parts: Study intro (10 minutes), Writing task (35 minutes), Interview (10 minutes), Questionnaire (5 minutes).}
\end{figure}

The study was conducted remotely using video call software and screen sharing. The sessions were scheduled for 60 minutes and structured as described next (\cref{fig:study_procedure}):

\subsubsection{Study Intro (10 minutes)}
An intro page on our web app explained the study in line with our institutional regulations (incl. GDPR, privacy and data collection info and informed consent).
A researcher gave a live demonstration of creating and using personas in our prototype, and encouraged participants to ask clarifying questions. %

\subsubsection{Writing Task (35 minutes)}
We asked participants to share their screens with recording. We explained the task of working on their text. During the recruitment, we had asked participants to think of writing tasks that they could bring to the study. Eleven people brought an actual writing project, 5 had come up with one for the study. %
A researcher observed the task and took notes. They were muted for most of the time, occasionally reminding participants to share their thoughts or asking a question to better understand people's thinking. 

\subsubsection{Interview (10 minutes)}
\label{sec:methods_interview}
We conducted semi-structured interviews after the writing task to further learn about people's experiences. %
We asked these main questions: ``What was your thinking when creating the personas?'',
``What did you think about the generated feedback?'',
``How did you use the feedback?'',
``What did you like about this idea?'',
``What could be improved/extended?''

\subsubsection{Questionnaire (5 minutes)}
Participants provided their demographics and in study 2 completed a final questionnaire asking about their experience with the overall concept as well as specific aspects and UI features. It included Likert items on \textit{Experience using the system} and \textit{AI interest}. We took the latter from the ``MeMo:KI -- Opinion Monitor AI''~\cite{memoki}. \revision{These Likert items were constructed as statements reflecting the participant's viewpoint, such as ``The feedback influenced my writing'', ``The feedback made it easier for me to write my text'', and ``I read articles about Artificial Intelligence with great curiosity''. See \cref{fig:likert_results} for all items.}

\subsection{Coding of Think-Aloud, Interviews, \revision{and the Generated Feedback}}\label{sec:methods_think_aloud}

The transcribed interviews and observation notes were analyzed using a German qualitative content analysis method~\cite{mayring2021qualitative} that is equivalent to thematic analysis~\cite{Braun2006,doi:10.1080/2159676X.2019.1628806}. Using axial coding principles~\cite{corbin2014basics}, the first author and second author individually reviewed the texts several times, moving back and forth through the material. They identified overarching categories/themes and subcategories of codes through an iterative process of clustering, splitting, and merging codes. The final themes were then discussed with the other authors in videoconferencing sessions. We repeated this process until a consensus was reached. \revision{In the same way, we also coded the generated feedback texts.}

%% file: content/05-evaluation.tex
\section{Results}
\label{sec:results}

\subsection{Overview} %

\subsubsection{Interaction (Logging Data)}
\begin{figure*}[t]
     \centering
     \begin{subfigure}[b]{0.27\textwidth}
         \centering
         \includegraphics[width=\textwidth]{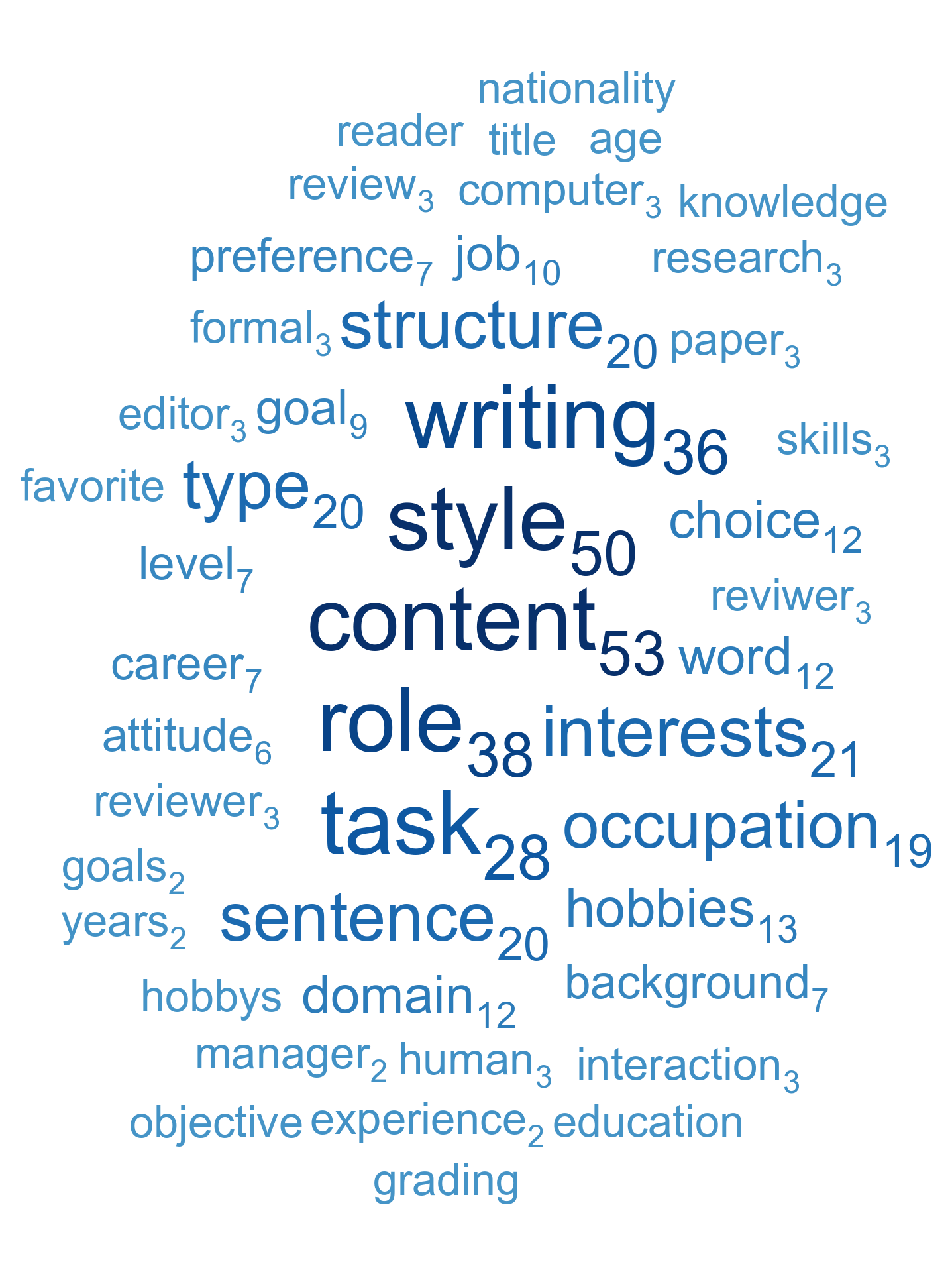}
         \caption{\revision{\textit{Attribute} words used in personas.}}
         \label{fig:wordcloud_attributes}
     \end{subfigure}
     \hfill
     \begin{subfigure}[b]{0.72\textwidth}
         \centering
         \includegraphics[width=\textwidth]{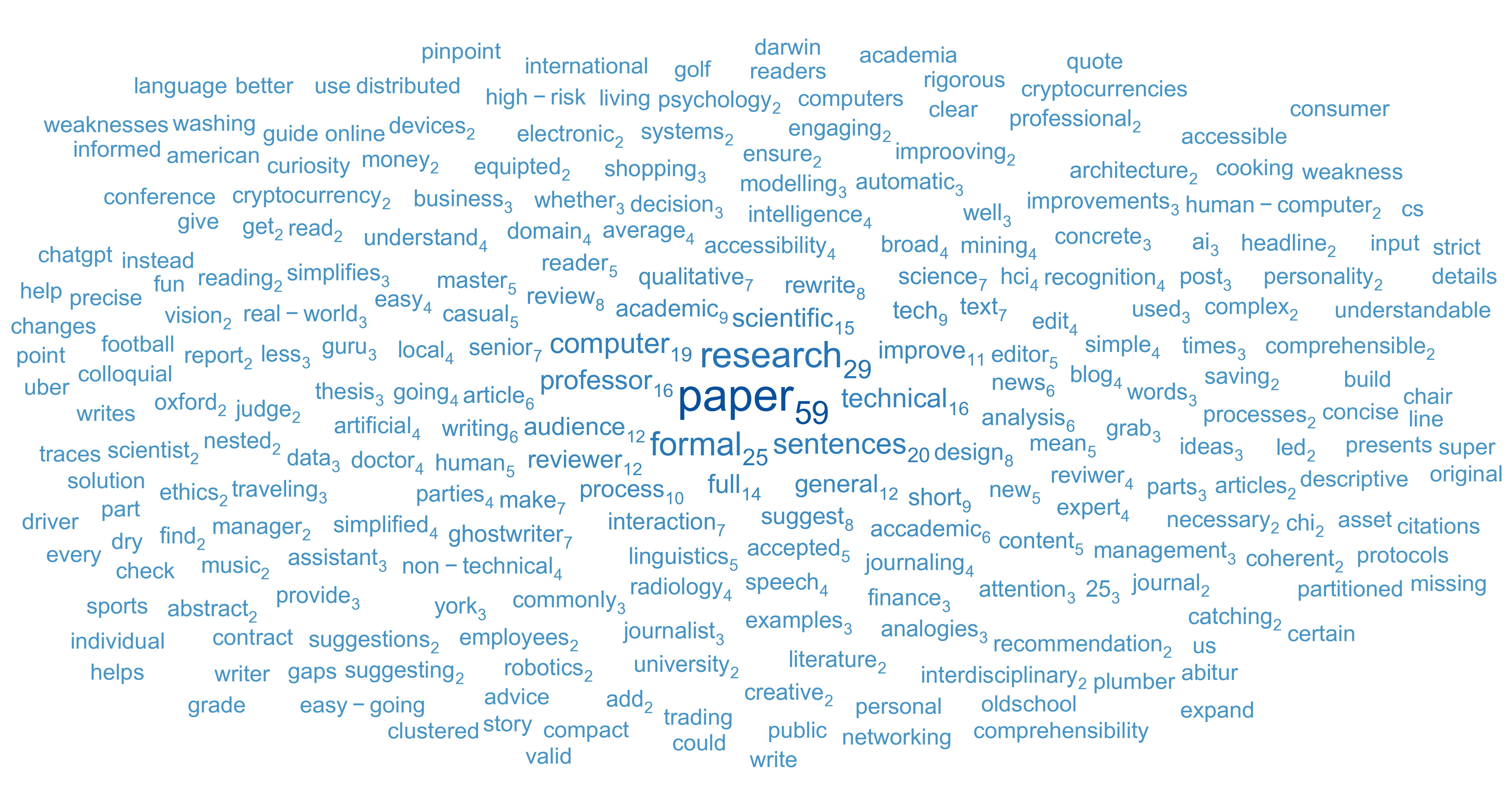}
         \caption{\revision{\textit{Description} words used in personas.}}
         \label{fig:wordcloud_values}
     \end{subfigure}
     \caption{\revision{Overview of the words that participants entered in the attribute-description pairs (see UI in \cref{fig:guidance}), lightly scaled by how often these contributed to generated feedback (also annotated as numbers if >1). For example, if a participant defined ``writing style: formal, scientific'' for a persona and requested feedback four times from this persona, this would contribute counts of four to ``writing'' and ``style'' (in \ref{fig:wordcloud_attributes}), as well as to ``formal'' and ``scientific'' (in \ref{fig:wordcloud_values}). We removed stopwords (e.g. ``the''). While shown as single words here for a better overview, participants also entered descriptions as phrases, not only single words (e.g. ``Provide real-world analogies to ideas in the paper''). Log files including verbatim descriptions are available in the project repository (see link in \cref{sec:conclusion}).}}
     \Description{Figure with two separate word clouds: The left one shows the words that participants entered as attributes of personas (top words: ``content'', ``writing'', ``style'', ``role'', ``task''). The right one shows the same for the words that participants entered as descriptions (top words: ``paper'', ``research'', ``formal'', ``sentences'', ``computer'').}
     \label{fig:wordclouds}
\end{figure*}

We logged interactions in study 2: 
On average, participants created 2.36 personas (SD 0.48, range 2 to 3). 
They requested a mean of 6.27 feedbacks (SD 2.83, range 3 to 11), and revisited created personas 7.46 times on average (SD 5.70). 
Their final texts in the editor had a mean length of 438 words (SD 284.14).
\revision{\cref{fig:wordclouds} shows an overview of the attribute-description pairs that participants entered to define their personas.}

\subsubsection{Perception of Interaction (Likert Data)}
\label{sec:likert_results}
Participants in study 2 rated several Likert items at the end (\cref{fig:likert_results}). In summary, these ratings overall indicate that participants found the feedback to be helpful and useful. %
Most rated the feedback as having impact on their writing, influencing content and/or wording.
Despite this influence, almost all felt that they were the authors of their text. They also did not feel as if they would be interacting with real people.
Participants were critical of the length of the feedback. Combined with the qualitative insights reported below, we learned that the feedback was often too long.

\subsubsection{\revision{Overview of the Generated Feedback}}
\label{sec:feedback_content_analysis}
\revision{
By manually coding and analyzing the 68 generated feedback texts (\cref{sec:methods_think_aloud}), we identified a recurring structure with the following building blocks:
In the \textbf{opening} sentences, the persona described its attributes, typically like this: ``As a [e.g., reviewer, professor, journalist, ...], with [certain attributes], my task is [...].''
This was almost always followed by a positive comment or summary of the user's selected text snippet, and often concluded with a suggestion on how to improve the text.}

\revision{
In the feedback's \textbf{main part}, the AI persona suggested adding more examples (47 occurrences), asked for topic-related content (26), sought clarifications (34) and more details (13), or gave specific hints, such as missing data, highlights, need for citations, or conclusions. In addition, the persona proposed improvements to the writing style (9), including suggestions for shorter sentences, easier language for accessibility, or different terminology.
This advice was typically first described on a higher level, followed by a concrete suggestion (27), most often with a variation of ``For example, instead of [text snippet], the author could write [proposed text].''}

\revision{
Finally, almost all feedback texts (66) concluded with a \textbf{summary}, often starting with ``Overall, the text snippet...'' or less frequently with phrases like ``In summary...'' or ``In conclusion...''.}

\revision{
Generally, we observed the AI personas to provide polite and uplifting feedback. 
Many times (63 occurrences), the feedback emphasized positive aspects of the text.
Even when one participant (\p{12}) created a ``mean'' persona, although less encouraging, the feedback remained polite and included self-critical statements like ``[...] the mean attitude in this context may affect the credibility of the feedback.''}

\revision{
While the feedback's structure consistently followed the described scheme, its content was tailored to each participant's text snippet and the persona's preferences and background, rather than offering generic suggestions (e.g. see~\cref{fig:interaction}).}

\subsection{How do writers perceive the idea of AI personas?~(RQ1)} 
\label{sec:results_rq1}

Participants used AI personas in various text types, including blog articles~(3), business reports~(1), research papers~(6), grant proposals~(1), research paper study notes~(1), essays~(2), and abstracts~(2).
Across all of them, they valued AI personas for gaining feedback from different perspectives. %
For example, \p{9} found it ``nice to create multidisciplinary personas to have different perspectives''. \p{12} valued ``to have, like, somebody, like, just write to you, OK, this needs more reasoning''. 
And \p{2} said that %
``it is important to know how a teacher who works at a school would go about this. What would her thoughts be?''

Eleven participants emphasized that it was a challenge to consider multiple points of view without AI personas, and saw the personas as a tool that helped them with that.
For instance, \p{13} said that ``it's often hard to think of these things on your own'' and ``to look at it from different angles''. Similarly, \p{4} compared the tool to ``a copilot to tell you, yeah, you forgot about this or that. That's good.''  And \p{6} described that the persona feedback ``[..] kind of makes you step outside of your writing'', including ``stuff I didn't notice before''.

\revision{Five people drew explicit comparisons to other AI tools.
For example}, \p{12} contrasted the concept with AI tools that give text suggestions: ``It's much more valuable than just rewriting the sentence, because for that I could just use, ChatGPT, for example.''. 
\revision{Similarly, \p{15} said: ``[the persona interaction] feels more like a person, like you kind of have this picture in your head of who you're talking to. It's a bit different with like generic AI models like ChatGPT [...].''}
\revision{When asked about what he liked about the concept, \p{13} highlighted that ``this layout and like being able to add personas [...] it's like a nice layer over whatever language model you're using underneath''. These comments point at the perceived usefulness of the persona concept beyond LLM-based support in general.}

\subsection{How do writers define and interact with AI personas?~(RQ2)}

We observed certain strategies that writers developed when interacting with the AI personas.

\subsubsection{Types and Roles of Personas}
\label{sec:results_persona_types}

Writers defined different types of personas and used them for different roles and goals. \cref{sec:appendix_persona_types} shows an overview.
A first type was expert personas. Participants defined experts -- %
such as History Professor, PhD Advisor, Doctor in Radiology -- to receive feedback, often on factual correctness. For example, \p{6} explained: ``I first defined a hardware expert to know what I was writing was correct and accurate''. %

Another type of personas was based on real individuals. %
For example, \p{14} said: ``I was actually imagining my professor that taught that course.'' %

A third type was based on roles in the publication process, such as ``Reviewer'' or ``Editor''. For instance, \p{1} said: ``Does this even fit with the publisher I'm applying to? You know, if I'm writing youth literature for children aged 10 to 14, then I can adapt that accordingly and look here.''

Another pattern was that four participants %
created personas for ``strict'' and critical perspectives on their text. For example, \p{12} said ``you wanna get the feedback from the mean reviewer''.

\subsubsection{Interacting with Personas}
\label{sec:results_interacting_with_personas}
Three patterns emerged for interaction with the personas (as opposed to the text): Initial exploration, defining personas, and refining personas.

\paragraph{Exploring AI Personas}
All participants found it interesting to explore the personas' capabilities. For instance, \p{8} said: ``I was super curious to see, like, how the different personas would give a different kind of feedback''. Sometimes participants %
were also just interested in exploring the concept through different kinds of feedback, such as \p{7}: ``Might be funny to define a persona like "mum"''.

\paragraph{Defining AI Personas}
Initially defining personas was challenging for participants, even after our design iteration added more guidance in the UI in V2 (\cref{sec:design_iteration}). %
For example, it was described as ``not that easy'' (\p{1}) and ``actually hard'' (\p{12}), and \p{16} had ``No idea, I'm not creative enough''. 
This was also evident in the time participants took to create their initial persona's profile. %
On average, participants (using V2) required 3.82 minutes (SD: 2.17), with the swiftest definition completed in just under one minute, while the longest duration exceeded seven minutes for a participant to define all attributes.

A strategy that participants used to overcome this challenge was to define AI personas based on real individuals, which makes this task more concrete. For example, this enabled \p{9} to use ``[that person's] backgrounds, their research interests, probably even the keywords that they have mostly on the papers''. %
Another pattern that helped them develop personas was iterative refinement, as described next.

\paragraph{Refining AI Personas}
Participants iteratively refined personas: They created a persona, requested feedback, and read that to decide whether it is the kind of feedback they wanted. If not, they added, removed or changed attributes. However, it was not always obvious how to achieve this. \p{10} wondered: ``What should I change? Is that level of granularity relevant?''. This workflow also motivated three participants to suggest conversations as a way to refine the persona, such as %
\p{12}: ``Can I iterate the feedback? I’m missing a chat here to refine. [...] It would be really cool to ask personas questions like \textit{would you use such a tool?}'' %

\subsubsection{Strategies for Writing with AI Personas}
\label{sec:results_strategies}

We observed different strategic workflows that participants employed when writing with the help of AI personas. While we describe them separately, they are not mutually exclusive.

\begin{figure}[t]
    \centering
    \includegraphics[width=\minof{\columnwidth}{0.8\textwidth}]{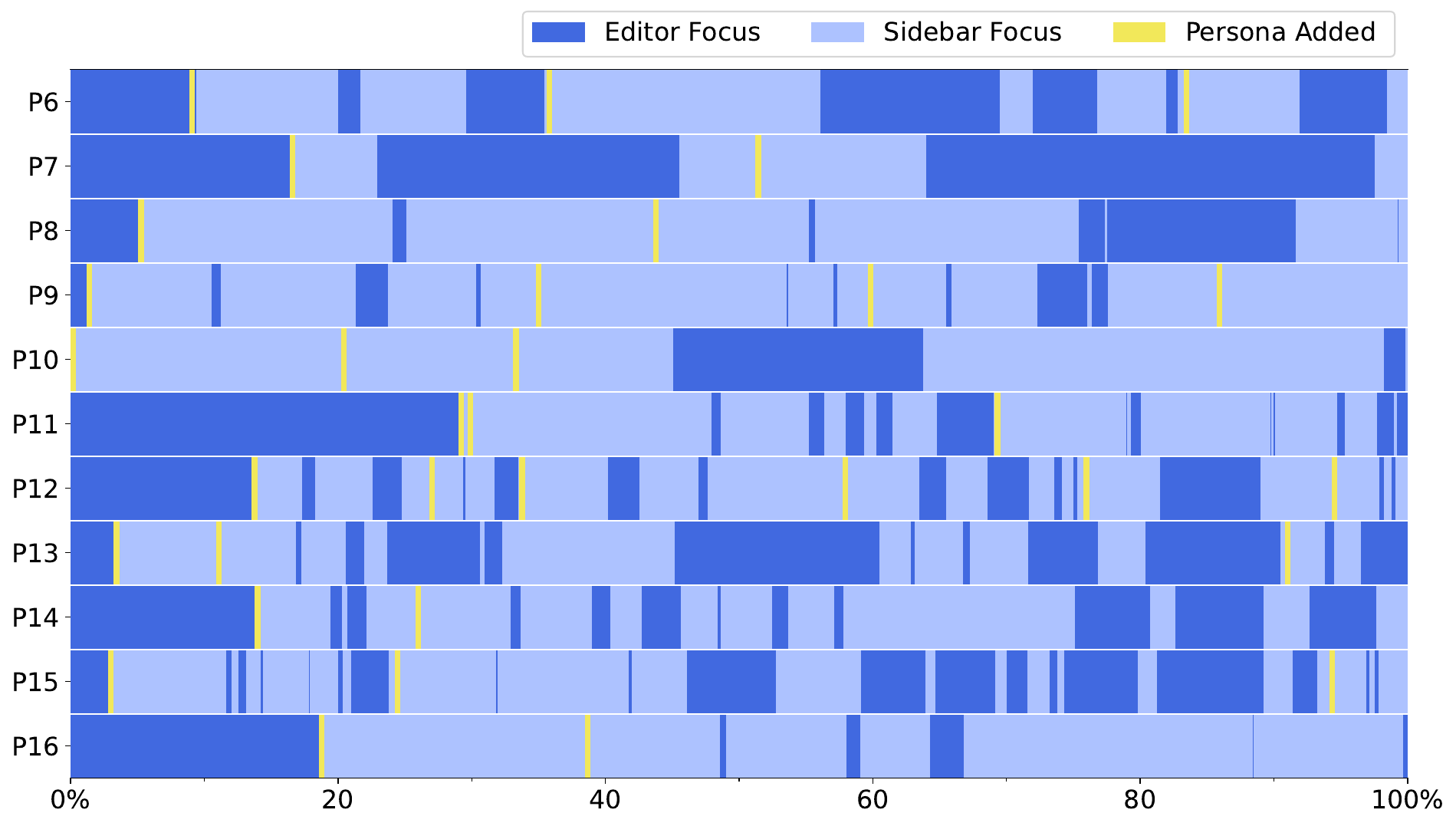}
    \caption{Overview of participants' workflows (in study 2): Each row (y-axis) is one participant. The x-axis shows time as study session progress. Color shows participants' focus on the editor (dark blued) or sidebar (light blue), as derived from logged UI events. Yellow lines indicate when a new persona was added. Overall, some participants switched between writing and engaging with personas more frequently than others. See \cref{sec:results_strategies} for details.}
    \label{fig:results_worflow_plot}
    \Description{Shows a vertical stack of horizontal bars, one for each participant. Each bar is colored in stripes that represent the participant's focus switches over time, between the text editor and the sidebar. Additional lines indicate the moments at which a new persona was added. The emerging patterns reveal that some participants switched between the two user interface parts more frequently than others.}
\end{figure}

\paragraph{Use of the Working Time}
\cref{fig:results_worflow_plot} shows how participants split their time between working in the editor and the sidebar (based on logged UI events): Some switched only a few times, focusing on either writing or engaging with the personas for longer periods. In contrast, others switched frequently. Across both these groups, most participants started in the text editor (e.g. copy-pasting in the writing project they brought to our study) and then switched to creating personas during the first third of the study.
\revision{
After engaging with their personas for the first time, participants asked for feedback approximately every 3 minutes and 37 seconds.
The most engaged participant sought feedback every two minutes, while the least engaged waited 4 minutes and 45 seconds.
}
\revision{
For participants who switched more frequently, two patterns emerge: Some (4) seek feedback mainly from various personas on the same text snippet, while others (2) consistently use a single persona across different text parts. %
}
\revision{We found no relationship between frequency of use and the perceived helpfulness of personas.}

\paragraph{Writing with AI Personas ``In-the-Loop''}
A key strategy that we observed among participants was an iterative ``feedback loop''. One example was \p{11}: After an initial feedback, they incorporated suggested changes into their text and requested another feedback from the same persona to verify whether they had improved the paragraph. We saw that writers were checking back with personas to see whether these were ``satisfied'' with the changes. For example, \p{8} said: ``Now I rewrite it myself and then I let the reviewer check it again and see if it actually made a difference''. %
And \p{10} said: ``It seems like the change is reflected in the new review''.

\paragraph{Writing with Multiple AI Personas}
All participants from both studies used multiple personas.
The key motivation was to gather different views. For example, \p{3} ``wanted it to be diverse'' and \p{13} wanted to ``simulate different reviewers''. A related strategy was to try and satisfy all these personas, by reflecting on and incorporating the feedback from all of them. %
Here, \p{6} assigned different personas to different sections -- a writing editor for their introduction with the task to reach a broader audience, and a hardware expert to check the main text. The Likert ratings also indicated that multiple personas were useful (\cref{fig:likert_results}).

\paragraph{Instructing AI Personas}
Participants were familiar with systems like ChatGPT.
We observed that four participants used prompting strategies from such systems to further steer the feedback:
For example, \p{15} adjusted the ``task'' attribute of their persona with the prompt ``find the original quote by [study author name]''. And \p{9} added an instruction prompt directly into the text for which they requested feedback (\textit{``Below is the discussion of a research paper being written and I need feedback based on the structure''}).

\subsubsection{Impact of the Generated Feedback}
\label{sec:results_feedback}

Participants found that the feedback helped them and had an impact on their text. We also observed impacts on the writers themselves.

\paragraph{Helpful Feedback}
All participants (16 people) explicitly commented on the feedback as being helpful. The positive perception is also reflected in the Likert ratings (\cref{sec:likert_results}). For instance, \p{6} said ``But once it actually got to the advice, I really, really liked it, like it was super helpful.''. The feedback was also deemed ``valid'' (\p{10}, \p{12}) and ``correct'' (\p{11}). 
It was also perceived as actionable: \p{11} used the feedback ``first, to actually review my writing process and second, to come up with new ideas'' and \p{14} mentioned that ``I would have incorporated more [feedback] if I [had the time]. They had or suggested to use further research and more studies and things like that, and I definitely would have done that.''
An indicator of believable feedback is that it reminded \p{10} ``of reviews [from real people] I have read in the past'' and \p{7} believed the real person ``would have answered pretty equal'' to their created persona.

To further improve the feedback, participants expressed their wishes for additional features, such as knowledge of related literature (\p{8}, \p{9}) or longer context (e.g., by allowing for multiple selections of text (\p{13}) or considering the overall document context (\p{9}, \p{16})).

\paragraph{Impact of the Feedback on the Text}
Most participants (12 people) made changes to their texts in our editor based on the feedback they received. For instance, \p{6} added a personal anecdote to their text after reading that the feedback suggested this. At the end, they said ``I ended up making a lot of adjustments based off the feedback.'' %
Related, 9 out of 11 participants in Study 2 reported in the Likert questions that the feedback influenced their writing (see \cref{sec:likert_results}).

Moreover, four participants %
either kept generated feedback or edits resulting from using our tool for further use beyond the study. For example, \p{12} copied one feedback into his own Overleaf project. And \p{4} worked on a grant proposal and found that ``this is actually quite useful, I will paste this paragraph into my proposal document. I can use it there.''

\paragraph{Impact of the Feedback on the Writer}
A different kind of impact was that for \p{6} feedback resulted in pressure of incorporating suggestions: ``I felt pressured to actually make the changes. Just because they're supposed to be like the expert in the scenario but it also felt strange to put their edits in exactly.''

Observations and Likert results indicate that writers were influenced in their writing. At the same time, the Likert results show that most participants (9 of 11 people in V2) also reported that they feel to be the authors of their text (\cref{sec:likert_results}). As \p{9} put it: ``I feel like I have been the author of my text. I do not that much for ChatGPT''.

Another finding was that we observed participants anthropomorphizing the system through their personas. For instance, \p{16} made a sad face after being told that the writing time was almost up, saying ``But I didn't talk to `best friend' yet'' before requesting feedback from that persona. Participants referenced their personas as if they could be real individuals. For example, they said ``Let's see what X thinks about this'', where X is the name of their persona (\p{6}, \p{12}). However, some participants (5 people) gave their personas generic names like ``Persona 1''. %

\subsection{How well can contemporary LLMs support AI personas?~(RQ3)}

Overall, we found that it is possible to use a contemporary LLM (GPT-3.5) to support writers with \aipersonas, albeit with some limitations and room for improvement, related to model and prompt engineering.

\subsubsection{Expectations about Contemporary LLMs}
As our participants were familiar with systems such as ChatGPT, some (3 people) stated that they recognize that the feedback was similar to that. For example, \p{13} said: ``Of course it looks a lot like something like ChatGPT''.
Related, some comments indicated that participants were interested in how their feedback is being generated. For example, \p{11} noted that ``It would be interesting to know what the data is it has been trained with''. The prototype was also explored by contrasting different personas, which points into the direction of inquiring into the system through interaction (cf. ``what if'', scrutability~\cite{Buschek2022howtosupport, Kay2013scrutability}).

\subsubsection{Quality of the Feedback}
As reported above (\cref{sec:results_feedback}), participants found the feedback helpful. The Likert ratings (\cref{sec:likert_results}) support this as well, with a majority perceiving it as helpful, influential, and making it easier to write their text. Together, these results show that the model we used (GPT-3.5), and the way we prompted it, was suitable to explore the concept of \aipersonas.

That said, participants raised three key issues about the quality of the feedback generation. Concretely, they criticized it as verbose, unspecific, and repetitive.

The feedback was longer than needed. For example, \p{8} said ``it took me a while to read like the actual feedback''. \p{10} and \p{15} suggested a ``bullet point'' format in contrast to the fully spelled out comments that were generated. 
Related to this verbosity, the feedback was partly perceived as unspecific, such as ``very generic'' (\p{9}) or ``too abstract'' (\p{10}). For example, \p{10} said that ``there was truth in there, but it was like a bit drowned out by the things that surrounded it''. And \p{11} said that ``It just tells me I have to do this and it doesn’t come up with the specific examples''. \p{12} wished for a ``follow-up feature where you can [...] ask it again, for example, just something specific''.

Closely related, participants perceived the feedback's pattern of connecting the suggestions to the persona's definition as repetitive. For example, \p{9} said that the first parts ``are a description [...] But the thing is, I wrote that. I know it's there. [...] I want just the last third''. Similarly, \p{10} said: ``I don't need them to tell me every time who they are.'' This perception is also reflected in the Likert ratings on suitable length (\cref{fig:likert_results}).

As a reminder, our rationale for prompting the system to respond in this way was to reveal to participants how their defined attributes informed the feedback. We return to potential improvements in the discussion (\cref{sec:discussion_prompt_engineering}).

\subsubsection{Controlling the Model}
Partly in response to the above deficits, several participants brought up ways of steering feedback through additional parameters that users set in the UI. \p{6} said that ``It would have been nice, like, if I could have kind of tuned the feedback I wanted. [...] I could have, like, highlighted it, but then also said, like, what type of feedback I want''. Similarly, to make feedback more specific, \p{9} proposed ``a third step'', where they can specify ``feedback based on the structure or based on the content''. Further feature suggestions included that the system should know about previous feedback, as well as giving personas instructions. %

%% file: content/06-discussion.tex
\section{Discussion}
\label{sec:discussion}

\subsection{Taxonomy of AI Personas for Feedback}

Based on our study, we provide a taxonomy of personas and which dimension users considered. The most frequently observed attribute that users took into account was \textbf{expertise}. In the specific academic setting we investigated, this was frequently related to competencies, professional roles and qualifications. %
Such expertise was also defined by mentioning specific, well-known institutions like Oxford University (\p{10}) or the New York Times (\p{13}). \p{15} used the abstract term ``scientific writer'', while \p{11} referred to a top researcher by name.

Moreover, \textbf{social relation} was an important consideration. This is reflected in personas named ``best friend'', ``my colleague'', and ``my manager''. One participant even mentioned that it might be fun to define a persona such as ``mom''~(\p{7}). 
We also observed that the \textbf{valence} of personas was taken into account. Some users considered a ``worst-case'' person like a ``mean reviewer''~(\p{12}), others emulated their ``PhD advisor''~(\p{14}) or their ``PhD colleague''~(\p{11}), as more positively supportive personas. 
Another dimension we identified was the \textbf{level of involvement}, again ranging from closely related personas, such as ``my manager''~(\p{7}) and ``my PhD colleague''~(\p{11}), to less involved personas such as a ``New York Times journalist''~(\p{13}).

The participants also specified certain characteristics of the personas, e.g., that the grading style of a professor is ``strict'' (\p{15}) or that reviewer's comments are ``critical'' or ``mean'' (\p{12}). In line with the kind of personas commonly used in design, some participants also mentioned the hobbies of a persona~(\p{11}).

Our study sample was mostly young academics. We expect that other populations will define different concrete personas. While the presented taxonomy is on a higher level of abstraction, insights from other user groups might extend or refine it. Future work could explore this further with different samples. \revision{The taxonomy may also have to be adapted for other writing styles. For instance, expertise may play a less important role in the context of fiction writing. Future work could explore how the different factors must be weighted based on genre.}

\subsection{The Benefits of On-Demand Feedback from AI Personas}
We identified three main aspects about \aipersonas{} that writers found valuable in our study.

\subsubsection{Writers value and use feedback from \aipersonas}
The \aipersonas{}  were useful for our participants. They were viewed as a ``cool concept'' (\p{7}) and ``super helpful'' (\p{6}) and recognized as providing different perspectives and helpful feedback. %
It is particular noteworthy that most participants either kept the feedback and the revised text for their projects beyond the study (\p{9}, \p{10}, \p{11}, \p{12}), or said they would have done so, if the study had been in line with their overall process for that work (\p{7}, \p{8}: if they had more time; and \p{13}, \p{14}, \p{15}, \p{16} had used past/published writing projects).
That is, even the feedback provided in the study will already have real-world implications.
In summary, we conclude that the concept and implementation were positively received. %

\subsubsection{Writers use \aipersonas{} to consider multiple perspectives}
One key benefit that we identified is that \aipersonas{} in our prototype empowered users to compare \textit{multiple} perspectives to reflect on their writing \revision{(cf. goal 4 in \cref{sec:design_goals})}. \p{6} reported several occasions where adding different perspectives made him ``step outside of [their] writing and look at it'', which helped them notice things they had not noticed before. \p{9} compared the AI persona approach to the video game Sims, where the user can ``create a lot of characters''. Strategically, writers added further personas throughout their writing time in the study (\cref{fig:results_worflow_plot}). Overall, participants created a diverse range of personas (\cref{sec:results_persona_types}, \cref{sec:appendix_persona_types}).

\subsubsection{Writers use \aipersonas{} to gain access to perspectives on demand}
The AI personas provided users with \textit{on-demand} feedback from the perspectives of people that they otherwise would not have been able to receive feedback from \revision{(cf. goal 1 in \cref{sec:design_goals})}. Participants liked the convenience of getting ``instant feedback''~(\p{14}). \p{13} also appreciated getting ``feedback from these different kinds of people without actually having to find them''. \p{7} described how they could create an avatar of professors that they have not met yet, based on a description of their backgrounds and research interests. While we rarely observed such advanced use of AI personas to ``simulate'' people a user has not yet met, it shows the potential of the concept and provides ample opportunities for future work.

For this discussion, we reflect on three specific examples of such on-demand feedback: The first example is \p{14}, who requested feedback from a professor who taught a particular course. This professor may have moved to a different university, and receiving feedback from them may not be possible anymore. Moreover, due to the many obligations of a professor, they may not have been able to provide feedback on demand. A second example is provided by \p{16}, who used the AI persona to receive feedback from a doctor in radiology. %
Radiologists are highly paid since they are scarce, in high demand, and their services are essential. The tool provided a radiologist's perspective on demand. The third example is \p{12}, who used the tool to receive feedback from ``the mean reviewer'' to improve the text. Reviewing scientific papers is a complex process that requires careful consideration. Accordingly, reviewers require sufficient time to write thoughtful reviews. With the AI personas, \p{12} received feedback from a reviewer perspective already during their work on the manuscript, before sending the paper to actual reviewers.

\subsection{The Challenges of Working with AI Personas for Feedback}
We identified two key challenges: Defining personas and verbose feedback.

\subsubsection{Writers find it difficult to define personas}
Initially defining a persona was challenging because it was difficult for participants to articulate the key aspects of the target group, \revision{even with the improved UI in study 2 (see \cref{sec:design_iteration} and goal 2 in \cref{sec:design_goals})}. Nevertheless, they found several strategies, such as thinking of a specific (real) person, using our provided example attributes (\cref{fig:guidance}), and iteratively refining a persona after reading its feedback \revision{(cf. goal 3 in \cref{sec:design_goals})}. Future research should explore further ways of supporting the creation of \aipersonas. For example, one idea is to provide a library of personas that allow writers to collaboratively refine and reuse personas. This is akin to prompt collections for LLM-based systems, where users share the instructions that they empirically found to be particularly helpful for a given task. We discuss another direction in \cref{sec:discussion_reader_defined_personas}.

\subsubsection{Generated feedback can be verbose, repetitive and unspecific}
The other key issue that we identified concerns text qualities. Participants found the feedback too verbose and -- connected to this -- repetitive and unspecific, \revision{although our UI enabled them to request  feedback for specific text passages, not the whole text (cf. goal 5 in \cref{sec:design_goals})}. One reason for this was that feedback started by repeating persona information. We discuss further prompt engineering as a solution strategy in \cref{sec:discussion_prompt_engineering}. Conceptually, based on participants' comments and iterative refinement of personas, we expect tradeoffs between feedback length and helpfulness as well as the transparency of the persona's ``argument'' (feedback is \textit{this} because of \textit{that} persona information). Future work could use our insights as a starting point to explore the ``sweet spot'' here. This might also be dynamic. For example, as some comments related to repetitiveness suggest, it might be useful to reduce feedback length over time as users become familiar with a persona.

\subsection{The Representativity of Contemporary AI Systems}
\label{sec:discussion_representativity}

\revision{As \aipersonas{} are a combination of user-defined personas and LLMs, they have the potential to reproduce harmful stereotypes. In our in-depth qualitative analysis, we did not identify any instances where participants perceived the feedback as perpetuating stereotypes. However, we identified situations where} they criticized that \revision{the system} unnecessarily repeated \revision{certain} attributes. While seeing how the LLM links personal attributes and feedback in its written output might be insightful (at least initially), it is also a likely place %
where stereotypes could surface.

\revision{Even though we did not encounter stereotypes in practice, the potential dangers of LLMs in this regard have been described in prior work, for instance, by \citet{10.1145/3442188.3445922}. Prior work provides} a broad overview of stereotypical and derogatory associations that LLMs infer from their training data, including sensitive attributes like gender, race, ethnicity, and disability status. Bender et al. argue that reliance on internet data is associated with a risk of encoding the dominant view and failing to represent changing social views after training (``value-lock''). They also explain why the mere size of the training corpus does not guarantee diversity since only a certain privileged group can fully participate online. This connects to work that discussed which human values are represented by LLMs and which ones are not~\cite{johnson2022ghost}.

These concerns are important for writing support systems since prior work showed that predictive keyboards~\cite{Arnold2020} and biased language models~\cite{10.1145/3544548.3581196} can influence people's text and attitudes. 
While not a major concern in our investigation, we found that \p{6} ``felt pressured to actually make the changes'', which indicates the potential power of feedback from \aipersonas{} in this regard. 

In view of these aspects, we consider it essential that the concept of \aipersonas{}, especially when put into practice, is pursued with careful consideration of how the resulting representations of people can be facilitated and evaluated. \revision{Our analysis of the generated feedback did not reveal derogatory or stereotypical statements, such as sexist or racist ones (cf. \cref{sec:feedback_content_analysis}). Nevertheless}, systems that provide \aipersonas{} could warn users of the potentially limited representativity of contemporary AI systems \revision{and how they can arise from limitations in the training data and the training processes}. 

One key part of our concept is that writers decide on what attributes to define for their personas, which they can change \revision{anytime}. While that is not the same as having control over the underlying LLM, it provides one pathway of influence to the user, which might be helpful in cases where writers themselves recognize generated output as unsuitable. 

\revision{That said, there might also be cases where the model's bias is more subtly expressed in the feedback, such that writers may not immediately notice this. Suppose stereotypes like a sexist statement are detected in generated comments in a research setting like our study. In that case, this may be addressed by sending an educational debrief to participants, as done in prior work~\cite{10.1145/3544548.3581196} (e.g., informing participants about the incorrect or biased statement, with a link/content from a credible source to counter it). In production systems, regular audits may be necessary to ensure that systems are as free of harmful stereotypes as possible~\cite{sandvig2014auditing}. For this, it may be helpful to inform users what stereotypes were detected, how they were detected, and how they were addressed.}

Looking ahead, we anticipate that these questions will not remain limited to the design of specific tools but likely impact HCI research more broadly. For example, it seems increasingly likely to receive AI-generated responses when conducting survey-based user research online~\cite{10.1145/3544548.3580688}.

\subsection{Implications for Research and Design}
\label{sec:discussion_implications}

Here we discuss implications for a shift in the design perspective of AI writing tools and two ideas for addressing the identified challenges in the future.

\subsubsection{Shifting the design focus: From human-AI to writer-reader relationships}

Recent related work is often framed around human-AI interaction (e.g. \cite{Dang2022, lee_coauthor_2022, Yang2020}). This emphasizes the relationship of humans and AI (or NLP~\cite{Yang2019}), where the human in focus of the design and research questions is the writer (cf.~\cite{Chang2023in2writing}). We propose an alternative framing -- designing for the \textit{writer-reader relationship}, which AI might support. This expanded view acknowledges two impacted human stakeholders explicitly. 

This reframing has implications for research and design, as we demonstrated in this paper, by taking the writer-reader relationship as a starting point for our concept development. Instead of asking how people might write with AI, we asked how AI might support people in writing for people. 

Related work on explainable AI (XAI) revealed interesting perspectives by recognizing XAI as ``socially-situated''~\cite{Ehsan2021exex}.
In our context, our socio-technical expansion of the design perspective on AI writing tools complements recent work on the social dynamics of writing support~\cite{Gero2023}, as well as calls for such tools to consider social aspects of language~\cite{kulkarni2023socialfactors}. It also contributes balance to design perspectives derived from the writing model by \citet{flower1981cognitive}, which is increasingly considered in the design of writing tools (see \cite{bhat2023interacting, Gero2022designspace}) but focuses on processes within the writer and thus has little to say by itself about the perspectives of other affected people, such as readers. 
Finally, we expect this framing to be fruitful beyond writing, for example, to also keep in mind the creator-audience relationship when designing tools for creators of images, videos, podcasts or games.

\subsubsection{Prompt engineering}
\label{sec:discussion_prompt_engineering}

The presented findings are based on one model and one way of prompting it. Thus, we do not claim to present the ``optimal'' implementation of \aipersonas{} for feedback. While further design iterations were beyond the scope of this paper,
\revision{we have started to explore two strategies to address verbosity, following the insights of our study: One strategy is to modify the prompt template to emphasize more strongly that feedback should be concise. In addition, we shortened the text in the few-shot examples. Anecdotally, first observations indicate that this may not be enough to keep generated comments concise consistently. As another strategy, we experimented with applying the model a second time to its own output, using a second prompt that instructs it to make the given feedback more concise. This seemed promising in our first observations and is in line with ``chaining'' in other prompt-based systems (cf.~\cite{Wu2022}).}

Future work could explore \revision{in detail how such} prompting techniques~ (e.g.~\cite{NEURIPS2020_1457c0d6, wei2022chain}) can be leveraged \revision{specifically} to make LLM output less verbose, 
\revision{since this is a broader issue beyond our use case: For example,} \citet{kabir2023answers} found LLMs to provide verbose answers to software engineering queries, \revision{and \citet{sun2023evaluating} diagnosed that LLMs struggle with hard constraints, including on length (``numerical planning'').} 
For our use case, further prompts might also be explored with regard to how representative the feedback is (see \cref{sec:discussion_representativity}). 

Underlying these considerations is the emerging question in HCI of how to support end users in prompting effectively~\cite{10.1145/3544548.3581388}. Here, the persona concept could be explored as a framing that allows non-experts to steer a model towards a desired role through an understandable UI and metaphor. In a way, this makes explicit that there are (customizable) UI concepts to design around the ``system instruction'' prompts of systems like ChatGPT (cf.~\cref{sec:background_nlp}). This direction also connects to recent work on configurable UIs for LLMs~\cite{Kim2023LMCanvas, Kim2023cells}.%

More broadly, with prompt engineering and LLMs, building functional prototypes of AI tools is becoming increasingly feasible in user-centred design. %
In this light, we see our work as a starting point towards understanding AI personas in various use cases. We hope others will join us to make them even more helpful for writers and other creators.

\subsubsection{Connecting writers and readers through reader-defined personas}
\label{sec:discussion_reader_defined_personas}

Given the complexity of defining a persona, it is essential to think of ways of supporting this further. One direction could be \textit{reader-defined personas}. Rather than asking individual writers to imagine personas that represent their readers, future work could ask readers to define their personas themselves. This would not only simplify the workflow for writers. Following the mantra ``Nothing About Us Without Us'', that scholars of Critical Disability Studies in HCI subscribe to~\cite{10.1145/3334480.3375150}, it would allow readers to articulate what they need and explicate what the AI-based persona should pay attention to. 
Conceptually, the AI would then serve both stakeholders (readers and writers) as a mediating ``executable proxy'' for communication: For example, the writer could refer to the AI persona in moments where direct communication with the readers is not possible, or use it to get feedback more frequently than what is otherwise feasible. %
Future work could explore this as an important kind of computer-supported cooperative work that socio-technical interventions can facilitate.

\section{Conclusion}
\label{sec:conclusion}

We have proposed the concept of \textit{\aipersonas{} for feedback}: %
Writers define personas of their target readers to receive feedback on their text. This feedback is generated by an LLM based on a prompt that includes the persona information. 

In our study, participants used these personas to get feedback from different perspectives but found it challenging to initially define them. The feedback was seen as helpful but too long. It inspired writers to make revisions that benefited the text in their view. %

We currently see a ``Cambrian explosion'' of LLM-based tools. %
Many use AI to generate text in lieu of human writing and lack a clear use case for doing so. 
Here, we challenge predominant design assumptions and goals by expanding the socio-technical perspective in the design of AI writing tools.
We call for recognizing the social nature of writing as a starting point for tool design. Complementary, we pursued a design goal that builds on the writer-reader relationship.

While AI feedback comes with limitations (e.g. lack of specificity, cf.~\cite{Gero2023}), it offers enticing benefits, such as direct editor integration, almost instant turnarounds, and infinite availability and patience with a struggling writer's repeated requests.
We look forward to future work towards the larger vision of leveraging AI to make us write better, to consider diverse reader perspectives, and to achieve a writing goal that is inherently human: sharing ideas, through language, that others can understand.

To facilitate such future work, we release our prototype and material here:

\url{https://osf.io/4awsm/}

%% file: content/07-appendix.tex
\section{Appendix}

\subsection{Prompt Template for Generating Feedback}
\label{sec:appendix_prompts}
This is the prompt template in Python that we used (some formatting added for the paper). The few-shot examples also followed this template and were inserted between the system description and the final instruction, as indicated below. This format is expected by the OpenAI API (Chat Completion).

{\scriptsize
\begin{lstlisting}

{
    "role": "system",
    "content": 'Personas are defined using four fixed attributes: role, background, style, and content. Each attribute consists of user-defined key-value pairs. The possible key-value pairs are not predefined and can vary. Generate persona-specific feedback for the text snippet highlighted by the user, considering the persona's unique attributes and any additional key-value pairs that might be defined by the user. You will take the role of the persona and write from their viewpoint. Every key-value attribute that is included in the personas definition describes the persona and therefore you. The feedback should align with the persona's characteristics and viewpoint, providing insights, suggestions, or comments that are relevant to the persona\'s role, background, style preferences, and content preferences.
    
    Input:
    Text: "Selected text snippet from the user\'s editor."
    Persona:
    - Role: {"key": "value"}
    - Background: {"key": "value"}
    - Style: {"key": "value"}
    - Content: {"key": "value"}
    
    Output: Generate persona-specific feedback for the provided text snippet based on the given persona attributes. Write the feedback as if you would be this persona. Consider the role, background, style, and content preferences of the persona. Provide insights, suggestions, or comments that align with the persona\'s characteristics and viewpoint. Feel free to incorporate any additional key-value pairs defined by the user in the persona definition to enhance the relevance of the feedback. Write one continuous feedback that is not longer than 200 words.',
},
...few-shot examples...
{
    "role": "user",
    "content": 'Input:
    Text: "'
    + <@\textcolor{teal}{selectedText}@>
    + '"
    Persona:
    - Role: '
    + str(<@\textcolor{teal}{roleAttributes}@>)
    + "
    - Background: "
    + str(<@\textcolor{teal}{backgroundAttributes}@>)
    + "
    - Style: "
    + str(<@\textcolor{teal}{styleAttributes}@>)
    + "
    - Content: "
    + str(<@\textcolor{teal}{contentAttributes}@>),
}
\end{lstlisting}
}

\revision{The attribute-description pairs defined by the writers were inserted as indicated by the Python statements above. For example, if a participant defined a persona with the pairs ``role: reviewer'', ``writing style: formal'', ``sentence length: short'', and ``occupation: CS professor'', and selected a text ``Lorem ipsum dolor sit amet'', then the corresponding part of the prompt template above would concretely look like this:}

{\scriptsize
\begin{lstlisting}
    Input:
    Text: "Lorem ipsum dolor sit amet"
    Persona:
    - Role: {"role": "reviewer"}
    - Background: {"occupation": "CS professor"}
    - Style: {"writing style": "formal", "sentence length": "short"}
    - Content: {}
\end{lstlisting}
}

\subsection{Feedback Interaction Example}
\label{sec:appendix_interaction}
\begin{figure*}[t]
    \centering
    \includegraphics[width=\linewidth]{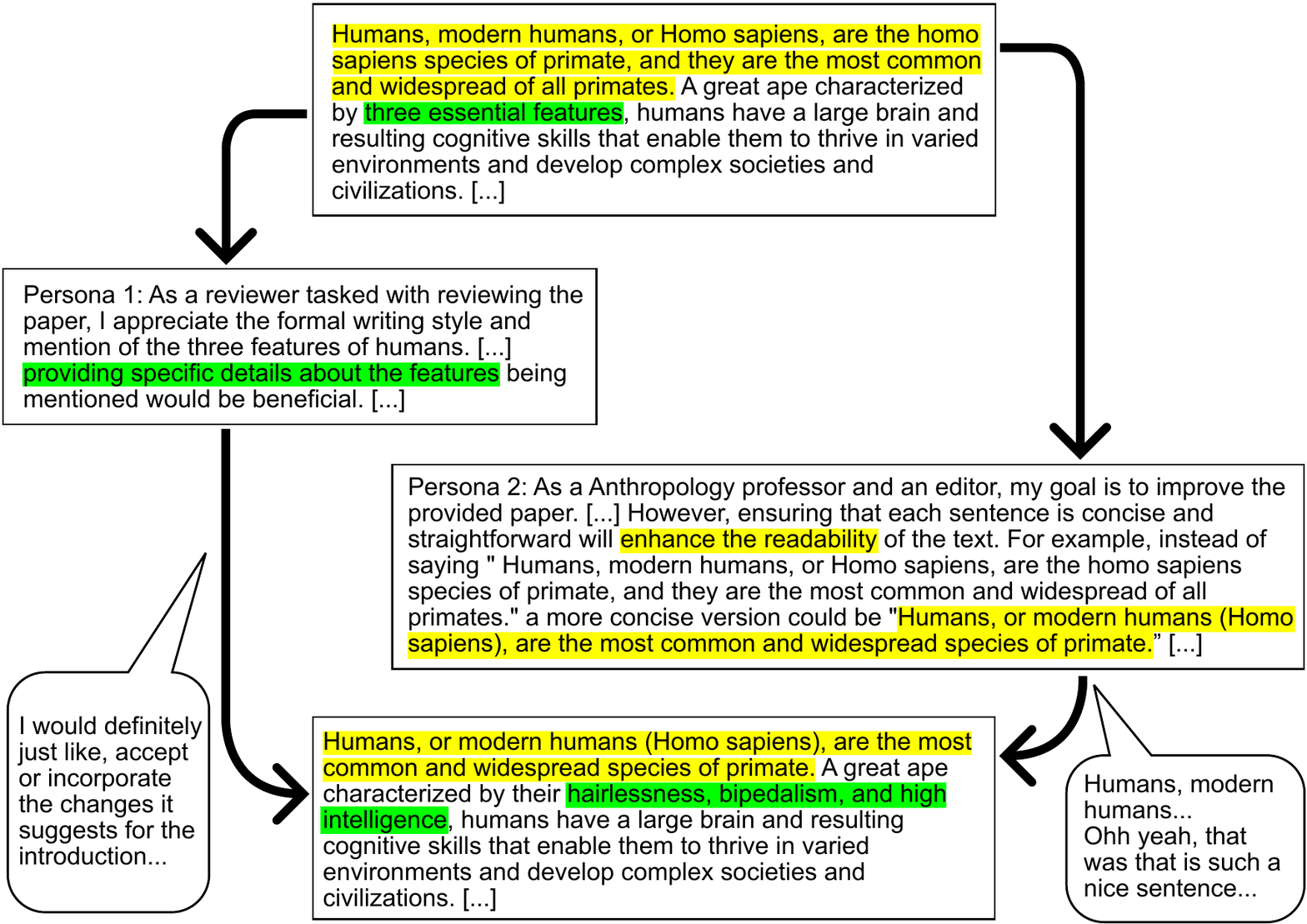}
    \caption{Example of a moment of interaction of \p{14} with two of their \aipersonas: \p{14} selected a paragraph in their draft (top box) and requested feedback from two personas (two boxes in the centre). The participant then made concrete edits based on both generated feedback (new draft version in the bottom box), and commented on this experience (speech bubbles). Ellipses ([...]) and color were used for this figure, not a part of how the text was shown in the UI. Note that we replaced the topic of the text in this example to preserve anonymity (see \cref{sec:appendix_interaction}).}
    \label{fig:interaction}
    \Description{This figure shows four texts: The original writer's text, feedback from Persona 1, feedback from Persona 2, and the resulting text after the writer made changes to their text based on the received feedback. Parts of the text that were impacted by the feedback are highlighted in color to show how the participant edited which parts of their text based on which parts of the feedback. In addition, the figure shows comments of the participant in speech bubbles.}
\end{figure*}

\cref{fig:interaction} shows an example of one concrete moment of interaction of one participant (\p{14}) with two of their \aipersonas. As shown, the participant selected a paragraph in their draft and subsequently requested feedback from two personas (a reviewer and an Anthropology professor acting as an editor). The participant made concrete edits based on both generated feedback and commented on this experience positively (speech bubbles in the figure). 

Note that we replaced the topic of the text in this example to preserve anonymity and avoid publishing the participant's text here. Concretely, we replaced the original topic and text with text from the Wikipedia entry on ``human''\footnote{\url{https://en.wikipedia.org/wiki/Human}}. The sentence structure, phrasing, semantics/scope of feedback and edits, etc. are all preserved as precisely as possible -- we only replaced the topic.

\subsection{Likert Results}
\label{sec:appendix_likerts}
\begin{figure*}[h]
    \centering
    \includegraphics[width=\linewidth]{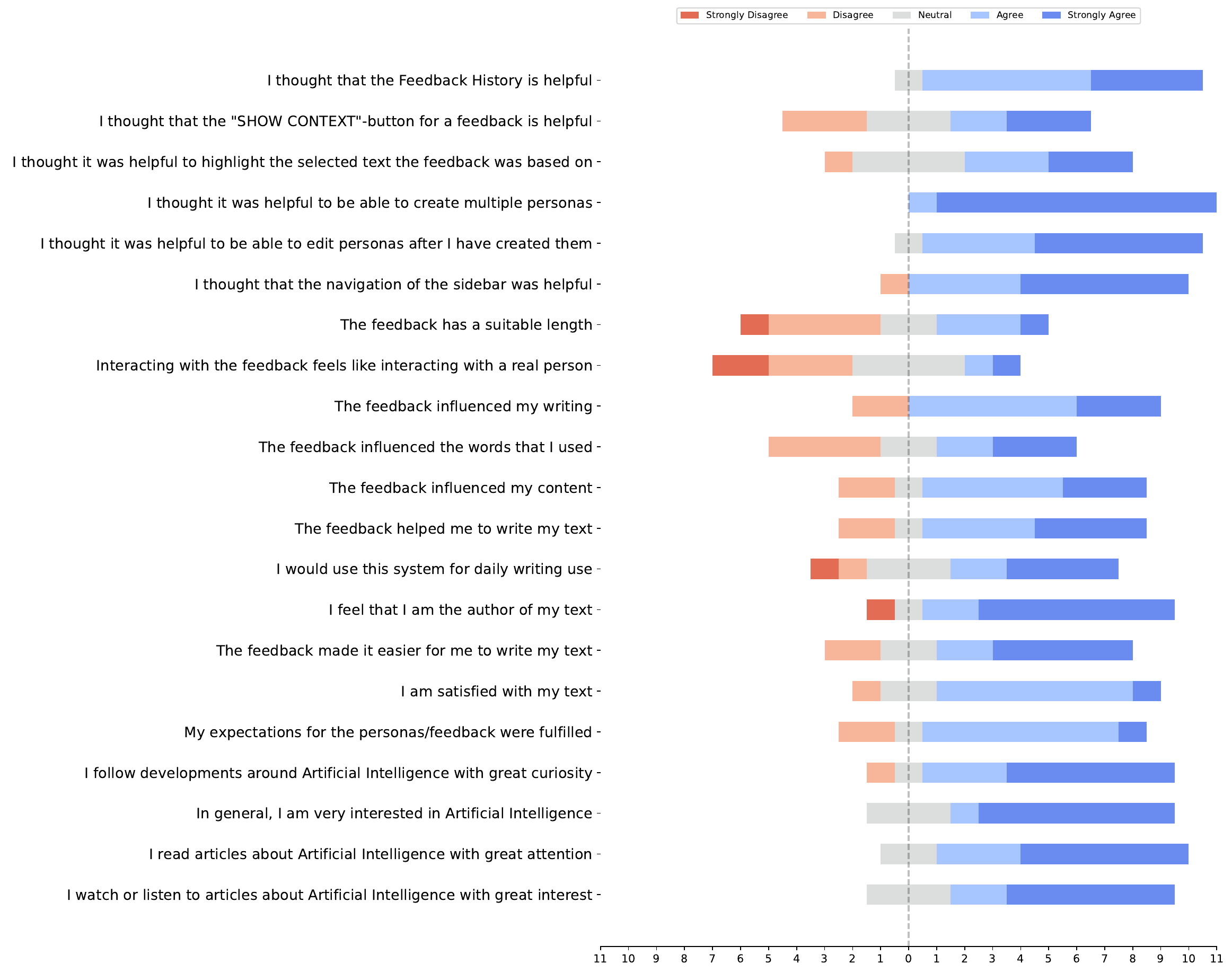}
    \caption{The Likert items on the concept and prototype, as well as interest in AI, asked at the end of study 2. Overall, participants appreciated the provided features. Most found that the feedback was helpful and had an influence on their work. They still perceived themselves as the authors. Participants were critical of the feedback length (too long). Finally, our sample had a high interest in AI.}
    \label{fig:likert_results}
    \Description{Shows the Likert data as horizontal diverging bar charts. Overall, participants appreciated the provided features. Most found that the feedback was helpful and had an influence on their work. They still perceived themselves as the authors. Participants were critical of the feedback length (too long). Finally, our sample had a high interest in Artificial Intelligence.}
\end{figure*}

\cref{fig:likert_results} shows the results from the final questionnaire from study 2. The items on interest in AI (bottom 4 items) were taken from the ``MeMo:KI -- Opinion Monitor AI''~\cite{memoki}.

\subsection{Persona Types}
\label{sec:appendix_persona_types}

\cref{tab:persona_types_s1} shows the different personas created in the first study, based on what participants entered for the persona (prototype V1 had a single text box, see \cref{sec:design_iteration}). For this overview, one researcher coded these and grouped them, resulting in the two shown dimensions (Gender, Age) with an additional ``note'' column containing further details. Prototype V1 had no dedicated name fields, so the three recorded names reflect how that participant referred to the persona in their verbal comments. 

\cref{tab:persona_types} shows the different personas created in the second study, where we logged the defined attributes and descriptions. For this overview, one researcher coded these and grouped them, resulting in the five shown dimensions ([Level of] Abstraction, [Writing] Style, Role, Focus [on], Background [related to text topic]) and their levels.
While the names are reported as entered by participants, the ``note'' column displays interesting additional information that was revealed by the participants during the study or in the following interview.

\begin{table*}[t]
\footnotesize
\begin{tabular}{lllll}
\toprule
P & Name        & Gender & Age         & Note                                   \\ \midrule
1 & Professor X & male   & -           & Professor for social gatherings        \\
1 & Frederik    & male   & 12          & Doesn't like reading                   \\
1 & John        & male   & -           & Persona occurs inside the narrative        \\
2 & -           & male   & 10          & -                                      \\
2 & -           & female & middle aged & housewife                              \\
2 & -           & -      & -           & school teacher                         \\
2 & -           & -      & middle aged & worker that wants to use AI            \\
2 & -           & -      & 60          & unfamiliar with chatbots               \\
2 & -           & -      & -           & teacher that wants to use AI           \\
3 & -           & -      & -           & specialized researcher                 \\
3 & -           & -      & -           & specialized reviewer from the US       \\
3 & -           & -      & -           & specialized reviewer from the UK       \\
3 & -           & female & -           & specialized reviewer from Australia    \\
3 & -           & male   & -           & specialized reviewer from India        \\
3 & -           & female & -           & specialized reviewer from India        \\
3 & -           & male   & -           & specialized reviewer from China        \\
3 & -           & female & -           & specialized reviewer from China        \\
3 & -           & female & -           & specialized reviewer from the UK       \\
3 & -           & male   & -           & specialized reviewer from the UK       \\
4 & -           & -      & -           & cybersecurity researcher specialized   \\
4 & -           & -      & -           & cybersecurity researcher unspecialized \\
4 & -           & -      & -           & politician with no technical knowledge \\
4 & -           & -      & -           & government worker for funding research \\
4 & -           & -      & -           & researcher evaluating novelty          \\
4 & -           & -      & -           & same researcher checking improvements  \\
5 & -           & -      & -           & professor                              \\
5 & -           & -      & -           & Italian chef                           \\
5 & -           & -      & -           & historian                              \\  \bottomrule
\end{tabular}
\caption{Overview of the different personas created in the first study (using prototype V1), based on what participants entered in the single persona description text box that was available in that prototype.}
\label{tab:persona_types_s1}
\Description{This table shows all personas that were defined in user study 1. Each persona has columns Participant, Persona Name, Gender, Age, and Note. One example is that participant 1 created a persona "Frederik" who is male, of age twelve, and the note explains that the persona doesn't like reading.}
\end{table*}

\begin{table*}[t]
\footnotesize
\begin{tabular}{llllllll}
\toprule
P  & Persona Name                  & Note                  & Abstraction & Style  & Role     & Focus & Background \\ \midrule
6  & Hardware Expert       & for fact checking     & medium      & formal & general  & -     & related   \\
6  & Writing Editor        & for accessibility     & high        & casual & general  & text  & unrelated \\
6  & Average Joe           & uninterested perspective   & medium      & casual & general  & text  & unrelated \\
7  & HR Manager Michael    & -                     & real person & formal & specific  & facts & related   \\
7  & Sybille               & participant's mom      & real person & formal & general  & -     & unrelated \\
8  & Crypto Bro            & perspective of reader & medium      & casual & general  & -     & related   \\
8  & Reviewer              & is a professor        & high        & formal & general  & -     & related   \\
9  & Persona 1             & is a HCI professor    & high        & formal & general  & -     & related   \\
9  & Persona 2             & writes solutions      & high        & formal & specific & -     & related   \\
10 & General Public        & educated but not expert    & medium      & formal & general  & text  & unrelated \\
10 & CS Professor, not HCI & Prof. in Computer Science (CS)          & low         & formal & general  & text  & unrelated \\
10 & CS Professor, HCI     & Oxford Prof. in CS and HCI  & low         & formal & general  & both  & related   \\
11 & [last name]           & famous professor     & real person & formal & specific & text  & related   \\
11 & Persona 2             & for rewriting text    & high        & formal & specific & facts & related   \\
12 & CHI Reviewer          & -                     & low         & formal & general  & -     & unrelated \\
12 & Researcher            & find good citations   & high        & -      & specific & -     & unrelated \\
13 & PhD Advisor           & modelled after real person & real person & formal & specific & facts & related   \\
13 & New to field          & NYTimes Journalist for simple text      & medium      & casual & general  & text  & unrelated \\
13 & Negative Reviewer     & worst-case persona    & medium      & -      & general  & both  & related   \\
14 & Persona 1             & -                     & high        & formal & general  & both  & -         \\
14 & Persona 2             & professor for grading & medium      & formal & general  & text  & related   \\
15 & Scientific Writer     & ghostwriter           & medium      & -      & specific & text  & unrelated \\
15 & Scientific Journalist & find specific information  & medium      & formal & specific & facts & unrelated \\
15 & Professor             & professor for grading & medium      & -      & general  & both  & -         \\
16 & Doctor                & expert in the field   & medium      & formal & general  & both  & related   \\
16 & Best friend           & for general support   & low         & casual & general  & -     & unrelated \\ \bottomrule
\end{tabular}
\caption{Overview of the different personas created in the second study (using prototype V2), based on the logged persona attributes and descriptions.}
\label{tab:persona_types}
\Description{This table shows all personas that were defined in user study 2. Each persona has columns Participant, Persona Name, Special Note, Abstraction Level, Writing Style, Role, Task Focus, and Background. One example is that participant 6 created a persona "Writing Editor" where the note explains that the persona was created for improving accessibility, with a high abstraction level and casual writing style. The role was general advice, the task was specifically focused on the text (not facts) and the personas background was related to the text's topic.}
\end{table*}